\documentclass[year=25]{fmcad}
\usepackage{amsmath,amsfonts}
\usepackage{algorithmic}
\usepackage{algorithm}
\usepackage{array}
\usepackage[caption=false,font=normalsize,labelfont=sf,textfont=sf]{subfig}
\usepackage{textcomp}
\usepackage{stfloats}
\usepackage{url}
\usepackage{verbatim}
\usepackage{graphicx}
\usepackage{cite}
\usepackage{tikz}
\usepackage{prettyref}
\newrefformat{li}{line~\ref{#1}}
\newrefformat{def}{Definition~\ref{#1}}
\newrefformat{eqn}{Equation~\ref{#1}}
\newrefformat{thm}{Theorem~\ref{#1}}
\newrefformat{lem}{Lemma~\ref{#1}}
\newrefformat{alg}{Algorithm~\ref{#1}}
\newrefformat{fig}{Figure~\ref{#1}}
\newrefformat{tab}{Table~\ref{#1}}
\newrefformat{app}{Appendix~\ref{#1}}
\newcommand\pref[1]{\prettyref{#1}}

\begin{document}

\title{Automated Formal Verification of a Software Fault Isolation System}

\author{
    \IEEEauthorblockN{
        Matthew Sotoudeh\IEEEauthorrefmark{1}\orcid{0000-0003-2060-1009},
        Zachary Yedidia\IEEEauthorrefmark{1}\orcid{0009-0006-4244-1690}
    }
    \\
    \IEEEauthorblockA{\IEEEauthorrefmark{1}Stanford University,
    Stanford, USA \\ \{sotoudeh, zyedidia\}@stanford.edu}
% IEEE Publication Technology,~\IEEEmembership{Staff,~IEEE,}
% % <-this % stops a space
}

\markboth{Journal of \LaTeX\ Class Files,~Vol.~14, No.~8, August~2021}%
{Shell \MakeLowercase{\textit{et al.}}: A Sample Article Using IEEEtran.cls for IEEE Journals}

\IEEEpubid{0000--0000/00\$00.00~\copyright~2021 IEEE}

\maketitle

\begin{abstract}
    Software fault isolation (SFI) is a popular way to sandbox untrusted
    software.
    A key component of SFI is the \emph{verifier} that checks the untrusted
    code is written in a subset of the machine language that guarantees it
    never reads or writes outside of a region of memory dedicated to the
    sandbox.
    Soundness bugs in the SFI verifier would break the SFI security model and
    allow the supposedly sandboxed code to read protected memory.
    In this paper, we address the concern of SFI verifier bugs by performing an
    automated formal verification of a recent SFI system called Lightweight
    Fault Isolation (LFI).
    In particular, we formally verify that programs accepted by the LFI
    verifier never read or write to memory outside of a designated sandbox
    region.
\end{abstract}


\section{Introduction}
Computer systems frequently need to run untrusted code, such as web browsers
running applets and scripts, software linking against third-party libraries, or
operating systems accepting user code into the kernel for packet filtering.
In all of these cases, security concerns demand that the host application only
run the untrusted code in ways that guarantee the untrusted code will not read
from or write to the host application's memory.
In theory, the untrusted code could be run in an entirely separate address
space (process) from the host code.
In practice, processes introduce far too much context switching overhead to be
reasonable for many applications.
Requiring the untrusted code to be written in a memory-safe language like Java,
JavaScript, EBPF, or WebAssembly that can be executed in the same address space
improves startup time but is significantly slower than native machine
code.
Furthermore, it makes the entire language interpreter or compiler a
security-critical part of the trusted codebase.

This paper focuses on the \emph{software fault isolation} (SFI) mechanism of
software sandboxing.
In SFI, the untrusted code is compiled to a memory-safe subset of the machine
code (ARM64).
That binary is only executed if it passes a \emph{SFI verifier}, which checks
that all memory access instructions in the binary target memory within the
sandbox.
Of course, verifying arbitrary machine code is hard; in practice, SFI
techniques require that the guest code be compiled to use a restricted subset
of memory operations that are easy to verify (a na\"ive SFI scheme, for
example, might require every memory read or write be immediately preceded by a
bounds check).

While SFI is a popular technique with many different
instantiations~\cite{nacl,veriwasm,lfi} we focus on the recent Lightweight
Fault Isolation (LFI) system~\cite{lfi}.
LFI is a very low-overhead sandboxing system specialized for ARM64.
There is ongoing work to bring LFI to production use-cases, and
large-scale deployment motivates extra scrutiny on the LFI verifier, which is
tasked with the security-critical job of verifying that a given binary cannot
access memory outside of its dedicated sandbox.

We discuss how the LFI verifier accomplishes this in~\pref{sec:Background}, but
briefly, it involves a whitelisted subset of ARM64 instructions that are chosen
to maintain certain memory isolation guarantees.
However, the ARM64 includes a very large number of instructions, and it is
conceivable that those in the whitelist interact in nontrivial ways to break
out of the sandbox.
Indeed, earlier versions of LFI have had soundness bugs because we did not
realize that certain whitelisted instructions could modify registers in
unexpected ways.

This paper describes our work on \textbf{automatic formal verification of the
lightweight fault isolation SFI system} for the ARM64 architecture; the LFI
verifier is now equipped with a formal proof that every program it accepts
cannot read outside of its dedicated sandbox memory region.

Our heavy automation is unique among existing formally verified SFI
systems~\cite{veriwasm}.
The only manual intervention needed is to write down an \emph{SFI invariant},
which requires only about 20 lines of SMT-LIB2 code.
To enable this, we codesigned the SFI system and verification pipeline.
Traditional SFI systems frequently check for the presence or absence of
\emph{sequences} of instructions and sometimes do complicated dataflow analyses
to ensure memory isolation.
But these nontrivial analyses are very difficult to formally prove
precise facts about.
Instead, we realized that the LFI scheme is essentially a \emph{stateless} SFI
system, i.e., its verifier is nothing more than an instruction whitelist.
The guest program may contain any instruction in the whitelist, in any order,
and no other instructions.
This means we need merely check that each whitelisted instruction, on its own,
maintains the expected SFI invariants.

The rest of this paper summarizes the LFI system~(\pref{sec:Background}),
describes our verification pipeline~(\pref{sec:Verification}), and summarizes
future and related work~(Sections~\ref{sec:Future}~and~\ref{sec:Related}).
An archival version of our source code and tools is available
in~\cite{artifact}.

% In any case, statelessness greatly simplifies the verification task.
% %
% In \pref{sec:Invariants} we define an \emph{SFI invariant} $I$ on CPU states.
% %
% Then, we use a formal model of the ARM instruction set to verify that if the CPU
% state initially satisfies $I$ and then executes a whitelisted instruction, it
% (1) still satisfies $I$ after the instruction execution and (2) does not
% perform any unpermitted memory reads or writes.
% 
% After defining the set of whitelisted instructions and the SFI invariant, the
% formal verification is \textbf{entirely automated}.
% %
% We iterate over the entire whitelisted subset of the ARM64 instruction set and
% use an SMT solver to check that the instruction maintains the SFI invariant and
% avoids unpermitted reads or writes.
% %
% To do so, we use ASLP~\cite{aslp} which is essentially a symbolic execution
% engine for the ARM Specification Language~\cite{asl} formalization of the ARM
% specification.
% %
% We parallelize the verification over a 190-core machine, and the entire
% verification takes about 20 hours to complete.
% 
% In summary, the contributions of this paper are:
% \begin{itemize}
%     \item A \emph{stateless} SFI system that enforces memory isolation using
%         nothing more than a per-instruction whitelist.
% 
%     \item An automated verification pipeline for stateless ARM64 SFI systems.
% 
%     \item The first automatically verified SFI system.
% \end{itemize}

%\clearpage
\section{Background on Lightweight Fault Isolation}
\label{sec:Background}
This section briefly describes the LFI system for software fault isolation.
Consider a setting where a web browser downloads and wants to execute
untrusted machine code from the Internet (similar to the NaCl model
used by Chromium).
In this setting, the website operator providing the machine code will compile
it using an untrusted, but LFI-aware compiler into ARM64 binary code.
This binary code is then sent to the browser, who wishes to run it while having
some guarantee that it will not perform unauthorized memory operations.
To do so, the browser provides the ARM64 binary to the \emph{LFI Verifier}, which
is essentially a sound program analyzer for memory isolation.
If the verifier rejects the program, it might be unsafe and so the browser
refuses to run it.
On the other hand, if the verifier accepts the program, it is guaranteed to be
memory isolated and so the browser can confidently run it.
The program might make calls into the LFI runtime to request actions like
drawing to the screen or sending network packets.

Safety relies on two crucial aspects: the LFI verifier and the LFI runtime.
A buggy verifier might accept code that reads from the host computer's memory
in insecure ways, while a buggy runtime might allow the binary to manipulate it
into performing unpermitted operations.
%
% Furthermore, the LFI verifier must assume that the LFI runtime is safe, i.e.,
% does not allow the binary to manipulate it into performing unpermitted
% operations.
%
% If the LFI runtime had a bug in it, the binary could ask it to perform unsafe
% operations even though the binary on its own is memory safe.

This paper considers only the LFI verifier, and so we assume the LFI runtime is
implemented correctly.
The notion of memory isolation here is more coarse grained than memory safety
in traditional programming languages: the untrusted program may access any part
of a 4GiB `sandbox' devoted to it.
It can do anything in that sandbox; it might interpret portions of it
as arrays and then index those application-interpreted arrays out-of-bounds, as
long as it never successfully\footnote{Notably our verification only shows that
the program never \emph{successfully} reads or writes outside of the sandbox;
valid LFI programs can sometimes make \emph{attempted} reads or writes as long
as those attempts are guaranteed to terminate execution, e.g., they access a
page that the runtime guarantees is unmapped. This is discussed further
in~\pref{sec:Redzones}.} reads or writes outside of its 4GiB sandbox.

The rest of this section briefly describes how the LFI verifier checks memory
dereferences in its simpler sparse mode; complete details on the LFI system are
available in~\cite{lfi}.
%
% We focus on e will describe how it verifies that load and store
% instructions all target the sandbox region; similar techniques verify
% correctness of the stack operations and indirect jumps.

\subsection{Reserved Registers for Memory Isolation}
The job of the LFI verifier (checking the given code is memory isolated) is, in
full generality, undecidable.
Hence, it must check instead for a subset of known-to-be-memory-safe constructs
(sound but incomplete).
% (i.e., always reject memory unsafe programs but sometimes fail to accept
% programs that are truly memory safe).
%
The set of constructs it checks for must allow efficient compiled code.

The primary trick is \emph{reserved registers}.
There is a reserved \emph{sandbox base register} (\texttt{x21} in LFI)
pointing to the first byte of the program's 4GiB sandbox region.
This region is assumed to be 4GiB aligned, so the bottom 32 bits of the
\texttt{x21} register are zero.
There is also a reserved \emph{addressing register} (\texttt{x18} in LFI) which
the verifier ensures is the only register (with some exceptions) used for the
ARM64 load and store instructions.
To a first approximation, the LFI verifier iterates over the binary,
disassembles each instruction, and checks:
\begin{itemize}
    \item No instruction ever writes to \texttt{x21}, so \texttt{x21} will
        always contain the sandbox base address.

    \item The only instructions that can write to \texttt{x18} are those of the
        form \texttt{add x18, x21, wN, uxtw}, which results in \texttt{x18}
        containing the high 32 bits of register \texttt{x21} concatenated to
        the low 32 bits of register \texttt{xN}.

    \item The only instructions that can load from or store to memory use
        \texttt{x18} as the memory address.
        % (These final two rules have some exceptions; see below for a
        % discussion of the stack pointer and~\cite{lfi} for a more complete
        % treatment.)
\end{itemize}
A program that passes this check is thus unable to explicitly load memory
outside of the 4GiB sandbox region, because it can only load from \texttt{x18}
and \texttt{x18} can only be set by an operation that ensures its final value
is within the sandbox region (i.e., has the same upper 32 bits as
\texttt{x21}).

Note that, beyond this simple approximation, there are many corner cases that
need to be handled~\cite{lfi}.
For example, \texttt{x18} might be set to the very last byte in the 4GiB
sandbox region, but then doing an 8 byte load from that address will actually
read outside of the 4GiB sandbox region.
To resolve this, the LFI specification~\cite{lfispec} requires that the sandbox
be surrounded by unmapped `redzone regions' so that such corner case reads
cause a trap rather than reading host memory.
Use of the stack pointer register presents other complications, described
below.

\subsection{Handling the Stack}
\label{sec:Redzones}
In addition to direct memory loads via the \texttt{x18} register, LFI also
allows loads relative to the stack pointer (\texttt{sp}).
This introduces extra complications as \emph{writeback loads} are used to
simulate a stack push.
For example, the instruction
\begin{verbatim}
    // sp = sandbox_base
    ldr x0, [sp], #-8
    // sp = sandbox_base - 8
\end{verbatim}
dereferences the stack pointer and then decrements it by 8 bytes.
But if the stack pointer is initially pointing at the base of the sandbox
region, after executing this instruction it will now be pointing outside of the
sandbox region.
To resolve this, the LFI runtime guarantees that 4GiB of address space
preceding and following the sandbox region is unmapped, so attempting to
actually access this region would cause a trap and hence safe exit of the guest
program (the largest offset accepted by the writeback load operation is smaller
than 4GiB).

The reader might be concerned that a chain of such instructions might leave
\texttt{sp} past even this 4GiB buffer region and hence possibly pointing at
host memory:
\begin{verbatim}
    // sp = sandbox_base
    ldr x0, [sp], #-8
    // sp = sandbox_base - 8
    ldr x0, [sp], #-8
    // sp = sandbox_base - 16
    ldr x0, [sp], #-8
    ...
    // sp = sandbox_base - 4GiB - 8?
    ldr x0, [sp], #-8 // load bad data?
\end{verbatim}
But indeed this is not possible because the second operation in this sequence
would trap (accessing unmapped memory), transfer control back to the LFI
runtime, and thus terminate the execution of the sandboxed code.

\subsection{Compiler Instrumentation to Generate LFI-Passing Code}
Compilers rarely produce code passing the LFI verifier, e.g., compiled code
often loads from a register other than \texttt{x18} or writes to \texttt{x18}
using a nonwhitelisted instruction.
Hence, the LFI project includes a modified compiler toolchain that
produces LFI-passing binary.
To do so, it tells the compiler to reserve registers \texttt{x18} and
\texttt{x21}.
Then, after the compiler produces an assembly file, we identify instructions
that would fail LFI verification and rewrite them into sequences of
instructions that pass LFI verification.
For example, loading from \texttt{x5} may not be safe because the LFI
verifier has no way of knowing that it points into the sandbox region:
\begin{verbatim}
    ldur x2, [x5]
\end{verbatim}
Instead, the rewriter might transform this into the two-operation sequence:
\begin{verbatim}
    add x18, x21, w5, uxtw
    ldur x2, [x18]
\end{verbatim}
Which replaces the upper 32 bits of \texttt{x5} with those of \texttt{x21}
before the load.
If \texttt{x5} was in-bounds at this point in the program execution,
then this transformation has no effect.
But if \texttt{x5} was out-of-bounds, this transformation causes
it to load an in-bounds address instead.
This may lead to an application-level bug in the untrusted code, but
effectively prevents it from accessing the host memory.

% \subsection{LFI Runtime}
% Finally, it is often desirable for sandboxed programs to have some way of
% communicating with the host, e.g., making network requests or producing output.
% %
% In LFI this is accomplished by runtime calls, where the sandboxed program can
% jump to a specified address that acts similar to a system call.
% %
% The runtime is also charged with setting up the sandbox for the process and
% ensuring that the registers are initialized properly (e.g., that \texttt{x21}
% does point to the sandbox base address).
% %
% Bugs in the LFI runtime are out-of-scope for this work, although there has been
% prior work investigating formally verified LFI runtimes~\cite{wave}.

\section{Verification of LFI}
\label{sec:Verification}
This section describes how we verify correctness of the LFI system.
Recall that the LFI system is a whitelist of allowed instructions and a set of
memory mappings guaranteed by the runtime (e.g., that the page following the
sandbox region is trap-on-execute).
There are actually two variants of LFI that we verify: sparse LFI (where every
sandbox is surrounded by 4GiB redzone regions of unmapped memory) and dense LFI
(where sandboxes are packed more tightly, with internal unmapped regions).
In this section we focus on verification of sparse LFI; dense LFI is verified
similarly.
In general, we verify the following:
\begin{quote}
    If the registers satisfy a certain \emph{SFI invariant $I$}, and the next
    instruction to be executed is one of the whitelisted instructions, then
    when executing the instruction both (1) no memory operations (reads,
    writes, executes) outside of the 4GiB sandbox succeed; and (2) either the
    execution terminates or the SFI invariant $I$ holds on the successor state
    as well.
\end{quote}
This is verified per instruction: so as long as the guest code starts in a
state satisfying the SFI invariant $I$ and all of the instructions in the
sandbox region satisfy that property, it will continually satisfy that
invariant during the execution of the guest program and not make any unsafe
memory operations until it terminates.
Execution can terminate by executing an instruction that causes a hardware trap
(like reading from memory that is unmapped or executing an undefined
instruction) or by calling in to the LFI runtime.

\subsection{Register Invariants}
In the sparse LFI variant, we verify the following invariants:
\begin{itemize}
    \item \texttt{x21} points to the special 4GiB-aligned sandbox base address,
        which does not change during execution.
    \item \texttt{x18} and the stack pointer are both in the range
        $[\mathtt{x21}-128\mathtt{MiB},
        \mathtt{x21}+4\mathtt{GiB}+128\mathtt{MiB})$.\footnote{In dense mode, this is $+/-8\mathtt{KiB}$.}
    \item The program counter (PC) is in the range $[\mathtt{x21},
        \mathtt{x21}+4\mathtt{GiB})$.
    \item $R_{30}$ is either in the range $[\mathtt{x21},
        \mathtt{x21}+4\mathtt{GiB})$, or contains one of three special LFI
        runtime call addresses.
\end{itemize}
Note the second of these is to handle the case of writeback load operations as
described in~\pref{sec:Redzones}.

\subsection{Detecting Bad Side Effects}
The main goal of our verification is to ensure that the guest never breaks out
of its sandbox.
We record any (symbolic) memory reads or writes performed during the (symbolic)
execution of the instruction.
We say a bad side effect occurs if any of them successfully accesses an address
outside of the 4GiB sandbox region (note that accesses guaranteed to fail are
not successful; e.g., reading from a page that the runtime guarantees is
unmapped terminates execution of the sandboxed program).
An instruction fails verification if it can ever perform a bad side effect when
starting in a CPU state that satisfies the invariants and expected memory
mappings.

\subsection{Memory Mapping Assumptions and Detecting Faults}
Per the LFI specification, we assume that the LFI runtime has set up the
following memory mappings:
\begin{itemize}
    \item The 4GiB regions before and after the sandbox 4GiB region are both
        unmapped, so accesses to them trap.
    \item The first 4KiB of the sandbox region is marked read-only, and the
        first three words at that location contain the addresses of special LFI
        runtime calls (analogous to system calls).
\end{itemize}

\subsection{Handling Faults}
We need to detect when the program execution terminates.
We call this a `fault,' and it can occur in one of a few ways:
\begin{itemize}
    \item The PC (before or after the instruction) is not 4 byte aligned or is
        in a region of memory that we know is mapped no-execute.
    \item A memory read or write occurs to an address that we know is unmapped.
    \item The PC becomes one of the three special trampoline addresses.
    \item An explicit trap occurs, e.g., an undefined instruction.
\end{itemize}

Some nuance arises here because, to the best of our knowledge, ARM does not
always make strict guarantees about the order of execution of traps.
Consider a write to a memory range crossing two bytes, $b$ and $b+1$.
If $b$ is unmapped memory but $b+1$ is host memory outside of the sandbox
region, the operation is safe only if the trap is guaranteed to occur
\emph{before} the $b+1$ write happens, i.e., if the hardware bounds checks
`eagerly.'
Rather than assume that the ordering of these checks is guaranteed or
deterministic across microarchitectures, we check a much stronger property: bad
side effects do not occur \emph{even if the traps always happen last.}

\subsection{Implementation and Formal Model of ARM64}
Critical to our approach is the ability to, given an instruction encoding and a
desired logical property about the CPU state before and after execution of the
instruction, check whether that property holds.
This requires a formal model of the CPU semantics.
For this, we relied on the ASL formal specification of ARM64~\cite{asl}.
In particular, we used the ASLP interpreter~\cite{aslp} which can partially
evaluate the ASL specification and returns a symbolic intermediate
representation that summarizes the effects of the instruction on a symbolic
model of the CPU.

We wrote a Python library that extracts this symbolic intermediate
representation from the ASLP interpreter into a usable Python form.
We then wrote a symbolic execution engine for this intermediate representation.
The output of the symbolic execution is a series of SMT-LIB2 assertions
encoding the execution of the instruction.
After adding assertions checking the SFI invariant, we use the \texttt{Yices2}
SMT solver to determine satisfiability~\cite{yices2}.

Writing the symbolic execution engine was simpler than it sounds, because we
only need it to be sound, not complete.
In particular, because so few of the operations actually affect the
reserved registers, we can leave many functions uninterpreted.

We ran into two issues with the ASL specification and ASLP partial evaluator
that required extra work.
First, the ASL spec and ASLP partial evaluator lacked proper support for
expressing atomic instructions, which led to internal ASLP exceptions when
partially evaluating those instructions.
To solve this, we commented out the atomics-related ASL code that was
causing the issues, as our verification is not interested in race
conditions.
Second, the ASL spec's handling of system access traps returns a special record
type which also causes internal ASLP errors.
We resolved this by replacing the system access trap generation functions with
an assignment to a special \texttt{SYMEX\_DID\_TRAP} variable that indicates to
our symbolic execution engine directly when a trap occurs.

\subsection{Parallel Verification and Results}
Our verification technique involves iterating through all whitelisted ARM64
instructions.
To do so, we first iterate through all of the ARM64 instruction set ($2^{32}$
total instructions) and check with the stateless LFI verifier for which ones
were accepted.
We evenly divide those accepted instructions across all of the verifier worker
threads.

We ran verification in parallel on a 160-core Ampere Altra ARM64 machine.
Verification for each of the sparse and dense LFI systems required
approximately 20 hours each.
Throughout the verification process we identified some inaccuracies and missing
details in the LFI specification~\cite{lfispec}, including how LFI runtime
calls are handled and memory mapping guarantees for the first and last
sandboxes.
We are in the process of correcting these oversights in a future version of the
specification, but these are only missing details in the written specification.
After clarifying these specification issues, the verification process completed and
found no bugs.

\section{Limitations and Future Work}
\label{sec:Future}
This section describes possible future work.
First, we could use our verification pipeline to automatically synthesize SFI
systems from their invariants by iterating over all $2^{32}$ ARM64 instructions
and including only those that maintain the invariants.
This could enable rapid experimentation with SFI systems, especially to
different domains like determinism.
%
% We are also interested in the feasibility of designing stateless SFI systems
% for other architectures (including RISC-V and x86-64), although compressed
% instructions complicate this.

Second, we could try to more fully automate our verification pipeline by using
invariant inference techniques to automatically derive the invariant for a
given SFI system.

To speed up our verification pipeline we could also check groups of
instructions at once, e.g., all add instructions targeting a nonreserved
register.
This would require the ability to extract shared semantics for groups of
instructions or instruction encodings with symbolic bits.
In theory the ASLP interpreter supports this~\cite{aslp} but we decided to
use the simpler strategy for this paper and found performance was reasonable.

This work did not consider safety of the LFI runtime, which remains an
unverified but trusted component of the system.
%
% Future work might investigate methods of proving that the runtime properly sets
% up memory mappings, reserved registers, etc.

In this work we focused on verifying an SFI system for the ARM64 architecture.
But LFI also supports the x86-64 architecture, and there has been recent work
attempting to extend it to RISC-V.
It would be interesting to produce a fully verified SFI system for x86-64 or
RISC-V.
We believe that the verification strategy in this paper would work well for
verifying the RISC-V version of LFI, which involves a similarly sized set of
whitelisted instructions and is also effectively stateless.
However, making the x86-64 LFI system stateless requires recognizing a number
of `macro-operations' that are actually sequences of other instructions
(similar to \texttt{nacljump} in the original NaCL paper~\cite{nacl}).
This leads to the set of whitelisted (macro)instructions for x86-64 LFI being
much larger than the set for ARM64 and RISC-V, making it infeasible to
enumerate through all of them.
In theory, as described earlier, a verification pipeline that lets us leave
some bits (e.g., immediate values) symbolic would make verifying x86-64 LFI
feasible as well.

\section{Related Work}
\label{sec:Related}
\paragraph{Software Fault Isolation}
Pittsfield was an early SFI system~\cite{pittsfield}.
The Native Client (NaCl) SFI system was shipped with Google Chrome and used for
website-loaded add-ons such as required (at the time) by Netflix~\cite{nacl}.
This shows the security critical position that SFI systems play, motivating the
use of formal verification to ensure the absence of bugs.
More recently, the LFI system has been proposed as an optimized SFI system for
ARM64.

To our knowledge, ours is the first stateless SFI system.
Native Client, for example, required the use of a special two-instruction
`nacljump' sequence of instructions to perform indirect jumps while maintaining
control flow integrity.
Automated verification of such a stateful SFI verifier would be more
complicated, because we can no longer look at each instruction on its own,
leading to an impractically large encoding space of instructions to
automatically iterate over.
Similarly, LFI as proposed in~\cite{lfi} allows some within-basic-block
optimizations, particuarly removing guards to stack pointer modification when
a stack access is performed later in the basic block.
Since then, we have found those optimizations can be removed with minimal
performance impact, making it fully stateless.

\paragraph{Formal Models of ISAs}
This work used the ARM64 formal specification extracted by the ASLP~\cite{aslp}
partial evaluator.
\cite{x86k}~and~\cite{sail} are other formal specifications of hardware ISAs.
At the beginning of this project we tried to use the SAIL~\cite{sail}
architecture semantics.
In theory, using SAIL would have made it easier to extend our technique to
architectures beyond ARM64, since SAIL contains models of multiple different
architectures.
However, we found that the SAIL models simulate the architecture at a lower
level than we needed.
Furthermore, because the SAIL models were so fine-grained, they required
significantly more investment for us to understand the model and even
significantly more time to just run the compilers necessary to work with the
model.
Because our verification technique involves running a symbolic execution of the
architecture semantics for every one of the LFI-whitelisted instructions, we
decided to instead go for the ASLP semantics which were both simpler for us to
work with and had faster tooling.

\paragraph{Verified Sandboxing}
While we believe ours is the first \textit{automatically} verified SFI system,
there have been formally verified sandboxing systems before.
VeriWasm~\cite{veriwasm} has an interactively verified SFI system used as a
WebAssembly target.
vWasm~\cite{vwasm} is a verified compiler from WebAssembly to x86-64 machine
code written in F$^\ast$.
WAVE~\cite{wave} verifies a WASM runtime.
RockSalt~\cite{rocksalt} and ARMor~\cite{armor} use interactive theorem provers
to verify classic SFI verifiers.

\paragraph{Verified Program Analyses}
The LFI verifier is itself essentially a program analyzer.
The problem of verifying program analyzers has been explored in prior work,
such as ALIVE~\cite{alive} that verifiers program analyzers and transformations
in a domain-specific language, and ALIVE2~\cite{alive2} that does bounded
translation validation for LLVM optimizations.

\section*{Acknowledgements}
We thank the National Science Foundation (DGE-1656518, DGE-2146755), Stanford
IOG Research Hub, and Stanford Future of Digital Currency Initiative for
supporting grants.

\bibliographystyle{IEEEtran}
\bibliography{main.bib}

\end{document}